# Can New Light Rail Reduce Personal Vehicle Carbon Emissions? A Before-After, Experimental-Control Evaluation in Los Angeles

Marlon G. Boarnet[a], Xize Wang[b,*], Douglas Houston[c]

**ABSTRACT**

This paper uses a before-after, experimental-control group method to evaluate the impacts of the newly-opened Expo light rail transit line in Los Angeles on personal vehicle GHG emissions. We applied the California Air Resources Board's EMFAC 2011 emission model to estimate the amount of daily average $CO_2$ emissions from personal vehicle travel for 160 households across two waves, before and after the light rail opened. The 160 households were part of an experimental – control group research design. Approximately half of the households live within ½ mile of new Expo light rail stations (the experimental group) and the balance of the sampled households live beyond ½ mile from Expo light rail stations (the control group). Households tracked odometer mileage for all household vehicles for seven days in two sample waves, before the Expo Line opened (fall, 2011) and after the Expo Line opened (fall, 2012). Our analysis indicates that opening the Expo Line had a statistically significant impact on average daily $CO_2$ emissions from motor vehicles. We found that the $CO_2$ emission of households who reside within ½ mile of an Expo Line station was 27.17 percent smaller than those living more than ½ mile from a station after the opening of the light rail, while no significant difference exists before the opening. A difference-in-difference model suggests that the opening of the Expo Line is associated with 3,145 grams less of household vehicle $CO_2$ emissions per day as a treatment effect. A sensitivity analysis indicates that the emission reduction effect is also present when the experimental group of households is redefined to be those living within a kilometer from the new light rail stations.

a. Price School of Public Policy, University of Southern California, Email: boarnet@usc.edu. OCRID: 0000-0002-0890-347X.
b. Price School of Public Policy, University of Southern California, Email: wangxize316@gmail.com. OCRID: 0000-0002-4861-6002.
c. School of Social Ecology, University of California, Irvine, Email: houston@uci.edu.
*. Corresponding author.

## 1. INTRODUCTION

The context for regional transportation and land use planning has changed dramatically in the past few years. Cities across the world are investing in alternatives to automobile travel, in part with hopes that such investments will reduce greenhouse gas (GHG) emissions in the ground transport sector. In this paper we report the results of a before-after, experimental-control group evaluation of travel changes coincident with the opening of a new light rail line in Los Angeles, California. Experimental evaluation is still uncommon in transportation studies, and our results are the first to our knowledge that use experimental evaluation to quantify personal vehicle emission reductions from travel changes that occur after a new light rail line opens.

Our context, Los Angeles, California, is an important policy setting. California State Senate Bill (SB) 375 requires that metropolitan planning organizations (MPOs) be able to credibly quantify the impact of transportation investments on GHG emissions. The targets have been set by the California Air Resources Board (ARB), and the greater Los Angeles region's first plan to comply with those targets has been developed (Southern California Association of Governments, 2012). As part of a broad program of transportation investment, Los Angeles is pursuing what is likely the most ambitious transit construction program currently underway in the United States. Eighty percent of the transportation sales tax revenues in Los Angeles County will fund either bus or rail transit, based on the earlier Propositions A and C (passed in 1980 and 1990, respectively), and the more recent Measure R (passed in 2008.) The Los Angeles Metropolitan Transportation Authority is planning to open six new rail transit lines between 2012 and 2020 (Los Angeles Metro, 2009), of which the Expo Line Phase I, the subject of this study, is the first. Recent studies such as the one by Su and DeSalvo (2008) find that subsidies for public transportation infrastructure are associated with metropolitan areas that are less sprawled and with residents that drive less. However, our current empirical knowledge about how investment in rail transit will impact driving, transit use and carbon emission is far from complete.

There is a longstanding debate about whether the association between land use variables and travel behavior is causal (e.g. Boarnet (2011); National Research Council (2009)). Residential selection is the threat to causality that is most commonly discussed in the literature. Persons may sort into neighborhoods by choosing to live in locations that support their desired travel patterns. For a review of the literature that seeks to econometrically address this issue in cross-sectional studies, see (Cao, Mohktarian, and Handy, 2009).

The research summarized here gives results of a before-after, experimental-control group method to evaluate the impacts of the newly-opened Expo light rail transit (LRT) line on personal vehicle GHG emissions of nearby households. Evolving new data collection technologies provide more and higher-quality data for transportation policy and regional science research (Miller, 2010). This progress makes program evaluation methods more promising. To date, though, before-after experimental-control group evaluation has only been rarely applied in travel behavior studies.

Two previous studies used experimental-control, before-after research designs to evaluate the effect of light rail transit on travel behavior, in Salt Lake City, Utah (Brown and Werner, 2008) and Charlotte, North Carolina (MacDonald et al., 2010). Additionally, Guo, Agrawal, and Dill (2011) applied an experimental-control group method to study the effect of a vehicle mileage pricing experiment, and pricing's interaction with land use, on driving in greater Portland, Oregon. To our knowledge, no study has applied a before-after, experimental-control method to study the impacts of light rail transit on personal vehicle GHG emissions. While VMT is an important determinant of personal vehicle emissions, we are able to match driving to the emissions characteristics of our study sample's household vehicles, and track any changes in vehicle holdings over time. Both of these allow a more direct estimate of emissions changes than is possible by simply focusing on VMT as a proxy for GHG emissions.

By applying an experimental-control group method, we illuminate how a broad range of transportation projects and regional policies could be similarly evaluated. Section 2 introduces the data for the Expo LRT study; Section 3 shows the methods used to estimate carbon dioxide emission levels with emission models; Section 4 explains the difference-in-difference methods used to study the effects of light rail transit on daily household-level personal vehicle carbon emissions. Section 5 presents the results of the analysis; Section 6 discusses the impact on transit-related $CO_2$ emissions and how such impacts affect our results and Section 7 concludes.

## 2. DATA

*The Expo Line study*

The Expo Line is a light rail line in the Los Angeles metropolitan area that extends south and west from downtown Los Angeles. Phase 1 of the line, opened on April 28, 2012 and runs 8.7 miles from downtown Los Angeles westward to Culver City, near the junction of the 405 and 10 Freeways.[1] Construction of Phase 2, which will extend the line into downtown Santa Monica, began during the summer of 2012 and is scheduled to be complete in 2015. Figure 1 shows the Phase 1 portion of the line and its location within the Los Angeles metropolitan area.

---

[1] Some of the background material about the Expo Line and the survey and study design in this section also appears, in more detailed form, in Spears, Boarnet, and Houston (2015).

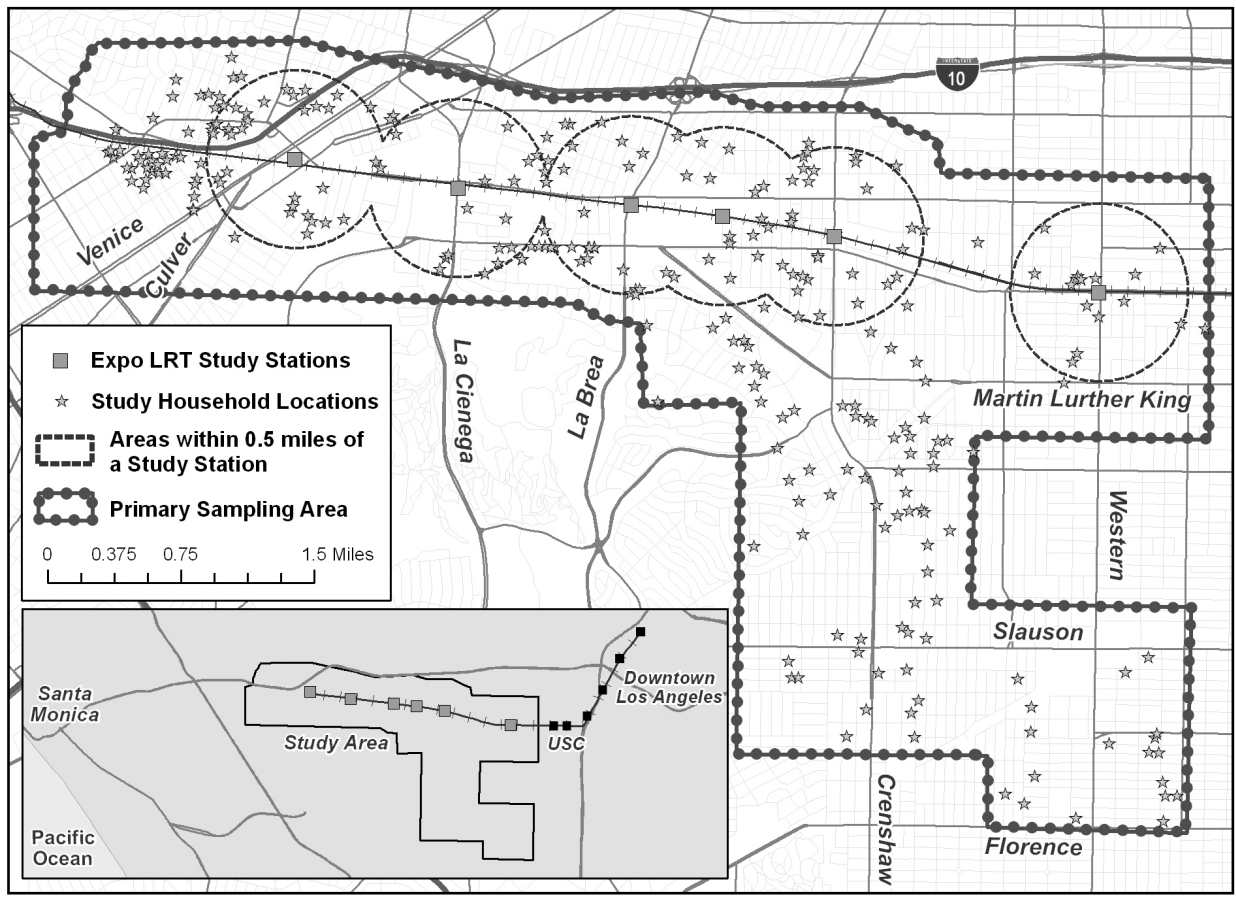

FIGURE 1: Expo Line Study Area and Experimental and Control Sampling Areas

Phase 1 of the Expo line stops at a total of 12 stations, ten of which were newly constructed. The analysis focused on residents residing near the six westernmost stations, and excludes impacts at the four easternmost stations near the University of Southern California (USC) since USC students living around these stations may be impacted by the LRT line very differently than long-term residents living in the study area.

In order to study the effect of the six westernmost Phase 1 stations on travel, we enrolled households into an experimental-control group study. Households in the experimental group are comprised of sampled residents living within half-mile radii of the stations. Control group households are comprised of sampled residents living beyond ½ mile from a station but in a neighborhood with built environment and socio-demographic characteristics that were similar to the experimental locations. The treatment and control areas are similar in terms of population density, age and income distribution. The approximate locations of the study households in the experimental and control areas are shown in Figure 1.

*The study sample*

We surveyed households in the experimental and control areas using paper and web-based questionnaires both before and after the opening of Phase 1 of the Expo LRT line. The survey asked households to log vehicle odometer readings for each of seven consecutive days for each household vehicle. The survey also asked a battery of socio-demographic and attitudinal questions and household members older than 12 years were asked to keep a count of their trips by mode for the seven day tracking period. Data collection for Wave 1 data, before the opening of the Expo Line Phase 1, was conducted from September 2011 to January 2012 with 284 households responding. Wave 2 data collection, after the opening of the Expo line, was conducted from September 2012 to January 2013 with 204 households out of the 284 in Wave 1 completing the second wave of data collection. The study also included a Wave 3, with 173 households responding. Due to sample attrition, for this paper we use only Wave 1 and Wave 2 data – one wave before and one wave after the opening of the line. See Spears, Boarnet, and Houston (2015) for evidence that travel behavior outcomes measured in Wave 2 generally persist into Wave 3 for these data, suggesting that a two-wave analysis for emissions is appropriate.

The study is a true panel; only households who completed a previous wave of data collection are invited into the subsequent wave of data collection. The respondents provided information about the make, model and year for each household vehicle. In order to maintain a balanced panel across the two waves, we only include the 160 households in both waves that have complete data to allow emissions calculations in each wave, as indicated in Table 1.

TABLE 1: Numbers of Households in the Survey

|  | Wave 1 | Wave 2 |
|---|---|---|
| Total number of households in the study | 284 | 204 |
| Total number of vehicles in the study | 354 | 249 |
| Number of households with matched data | 160 | 160 |
| Number of vehicles in households with matched data | 188 | 184 |

A detailed list of households and the criteria for keeping households in the balanced panel (both waves) is in Table 2. We dropped 124 households in wave 1 and 44 households in wave 2. The reasons why we dropped these households include: (1) the make, model or year of at least one vehicle in a household is missing; (2) the VMT for at least one vehicle in a household is missing, unreliable or an outlier (more than 200 miles per day on average over the 7-day survey period); (3) at least one vehicle of a household is either too old or too new, such that emission models do not contain sufficient information to estimate $CO_2$ emissions; (4) less than three days' odometer readings are available. If at least one of the conditions in (1) – (4) exists for a household in any wave, that household was dropped from the analysis. Of course, the 80 households that completed the Wave 1 data collection but which did not choose to participate in Wave 2 data collection are not included in the balanced panel data analysis reported below.

TABLE 2: Information of the Eliminated Households

|       | Missing/Incomplete Vehicle Info | Incomplete VMT info | No available emission factor | Unmatched | Total |
|-------|---|---|---|---|---|
| Wave1 | 27 | 35 | 2 | 60 | 124 |
| Wave2 | 8  | 7  | 3 | 26 | 44  |

## 3. ESTIMATING CARBON EMISSIONS

The estimation of emissions for gasoline vehicles is based on the EMFAC 2011 regional emission model developed by the California ARB (2011). Because the EMFAC 2011 model does not include hybrid or electric vehicles, the estimation of the $CO_2$ emission for hybrid or electric vehicles is based on the Fuel Economy online database developed by US Environmental Protection Agency (EPA, (2013). There are eight and nine hybrid or electric vehicles in Wave 1 and Wave 2, respectively, which is less than five percent of the total number of vehicles in either of the two waves.

For the gasoline fuel automobiles, the formula to estimate the daily average $CO_2$ emission of each vehicle is shown below:

(1) $\text{Daily average } CO_2 \text{ emission} = \text{Run rate (per mile)} \times \text{Daily average VMT} + \text{Start rate (per day)}.$

The "run rate" is the mass of $CO_2$ emitted by the vehicle per mile while running, the "start rate" is the mass of $CO_2$ emitted by the vehicle when the engine is turned on. Both rates, corresponding to the specific vehicle type (light-duty automobile, tier-1 and tier-2 light duty trucks, and motorcycles) and model year, are the average estimates in Los Angeles County during the year of the survey (2011 for Wave 1 and 2012 for Wave 2) available in the EMFAC 2011 emission database (California ARB, 2011). The survey queried trips for individuals but not for vehicles, hence it is not possible to infer the number of times each vehicle was started. We use the average start rate per day (rather than per trip) as an approximation, assuming that the number of trips for each vehicle is the average of the vehicles in the same type and model year in Los Angeles County. Both of the start and run $CO_2$ emission rates are the Pavley I - LCFS adjusted values based on California emission standards[2].

As to the vehicle type classifications, light duty automobiles (LDA) refer to regular-sized cars. The categorization of light duty trucks (LDTs), either tier-1 or tier-2, depends on the weight of the LDT vehicle.[3] Motorcycles (MCY) are also counted as an emission producer and included in the estimation.

For the small proportion of hybrid and electric vehicles, the formula to estimate the daily average $CO_2$ emission of each vehicle is as shown below:

$$(2) \quad Daily\ average\ CO_2\ emission\ = (tailpipe + upstream\ emission\ rate) \times daily\ average\ VMT.$$

The tailpipe emission indicates the mass of the $CO_2$ emission directly from the vehicle, while the upstream emission indicates the mass of the $CO_2$ emission from the power grids based on the electricity consumed. The emission rate per mile with these two sources combined is available for each specific vehicle type (make, model and year) from the EPA fuel economy online database (US EPA, 2013). Also, we assume a zero start rate of all hybrid and electrical vehicles given the lack of such information in the EPA database.

---

[2] The California Assembly Bill 1493, passed in 2002, authorized the California ARB to enforce the Pavley Clean Air Standards, beginning in 2009. For more details, see: http://www.arb.ca.gov/cc/ccms/ccms.htm.
[3] The standards are available at http://www.arb.ca.gov/msei/vehicle-categories.xlsx. The equivalent test weight is calculated as curb weight plus 300 pounds (see: http://www.arb.ca.gov/regact/2012/leviiighg2012/levappo.pdf). The curb weight data of the majority of the LDTs in the sample comes from "edmunds.com". A very small proportion of the curb weight information comes from Wikipedia for vehicles with no sufficient information in "edmunds.com".

After getting the daily average $CO_2$ emission for each vehicle, based on the emission rate and the average daily VMT, we can then calculate the household-level daily $CO_2$ emission levels (in grams) from personal vehicles for each household in each wave. One potential problem with this calculation is that the $CO_2$ emissions from alternative-fuel vehicles are based on the federal average emission standard, which is less strict compared to the California standard. However, we note that only a very small proportion of such vehicles exist (less than 5 percent) in both waves. Basic descriptive statistics of the household level daily average $CO_2$ emission levels are in Table 3 below. Table 3 also gives information on household travel data from the vehicle odometer and trip logs.

TABLE 3: Descriptive Statistics of $CO_2$ Emissions of Households in Both Waves

|  | Wave 1 | | | Wave 2 | | |
|---|---|---|---|---|---|---|
|  | all | control | experimental | all | control | experimental |
| Total number of households | 160 | 80 | 80 | 160 | 80 | 80 |
| Total number of vehicles | 188 | 99 | 89 | 184 | 93 | 91 |
| Average daily household-level VMT | 24.95 | 25.60 | 24.29 | 24.15 | 27.56 | 20.75 |
| Average daily household-level car trips | 4.51 | 4.71 | 4.32 | 4.16 | 4.46 | 3.87 |
| Average daily household-level bus trips | 0.46 | 0.50 | 0.42 | 0.47 | 0.57 | 0.38 |
| Average daily household-level train trips | 0.03 | 0.05 | 0.02 | 0.14 | 0.11 | 0.17 |
| Average of daily household-level CO2 emissions (grams) | 9681.9 | 9992.7 | 9371.1 | 9346.7 | 10815.9 | 7877.5 |
| S.D. of daily household-level CO2 emissions (grams) | 8698.3 | 7834.2 | 9524.1 | 9313.7 | 11129.9 | 6807.5 |
| Max. of daily household-level CO2 emissions (grams) | 56699.3 | 33615.5 | 56699.3 | 59660.6 | 59660.6 | 26935 |
| Min. of daily household-level CO2 emissions (grams) | 0 | 0 | 0 | 0 | 0 | 0 |

Half of the 160 households belong to the experimental group, while the remaining half are in the control group. The number of vehicles in the two groups is comparable for both waves, although the households in the control group have slightly more vehicles than the ones in the experimental group. The average daily household-level $CO_2$ emission varies from 0 to around 60,000 grams, while the experimental group in Wave 1 and the control group in Wave 2 have higher maximum emission levels. Note that the data already reflect an outlier removal process.

4. **ESTIMATION METHODS**

We compared the household-level personal vehicle $CO_2$ emissions for the experimental and control groups for each wave using two sample *t*-tests, and we also estimated the change across Wave 1 and Wave 2 for each group, the experimental households and the control households, using paired *t*-tests. In addition, we applied a difference-in-difference (DID) analysis to estimate the treatment effect of light rail line.

Defining $\mu_{it}$ as the mean of the outcome for group $i$ at time $t$, the DID estimator is $(\mu_{11} - \mu_{10}) - (\mu_{01} - \mu_{00})$, where $i = 1$ for experimental households and $i = 0$ for control households. This estimator can be evaluated using the following regression model:

(3) $$y_{it} = \beta_0 + \beta_1 X_i + \beta_2 T_t + \beta_3 X_i T_t + \mathbf{Z'}_{it}\boldsymbol{\beta_4} + \varepsilon_{it},$$

where $y_{it}$ is the daily personal vehicle $CO_2$ emission for household $i$ at time $t$, $X_i$ is a dummy variable where 0 represents the control group and 1 the experimental group, and $T_t$ is a dummy variable that takes the value 0 in the "before opening" period and 1 for the "after opening" period. The coefficient $\beta_3$ on the interaction variable $X_i T_t$ represents the DID estimator. Note that $X_i T_t$ takes a value of 1 only for experimental households in the after opening time period. The coefficient on $\beta_3$ is the effect of the treatment (in this case, the opening of the Expo LRT line) on the outcome variable among the experimental group (Card and Krueger, 1994). The vector $\mathbf{Z}_{it}$ indicates a vector of the sociodemographic variables used as controls in the model. The descriptive statistics for the dependent and independent variables of this DID model are in Table 4.

TABLE 4: Descriptive Statistics for Variables in the DID Regression

| Variable | description | Obs | Mean | Std. Dev. | Min | Max |
|---|---|---|---|---|---|---|
| emi | Daily average household-level $CO_2$ emission in grams | 320 | 9514.26 | 8998.68 | 0 | 59660.59 |
| veh_cnt | Number of household vehicles | 320 | 1.16 | 0.67 | 0 | 3 |
| ppl_cnt | Household size | 320 | 1.64 | 0.83 | 1 | 5 |
| experimental | =1 if in experimental group | 320 | 0.50 | 0.50 | 0 | 1 |
| wave | =1 if in wave 2 | 320 | 0.50 | 0.50 | 0 | 1 |
| wvexp | experiment * wave interaction | 320 | 0.25 | 0.43 | 0 | 1 |
| HHINC=1 | HH income less than $15,000/yr | 42* | - | - | - | - |
| HHINC=2 | HH income $15,001 to $35,000/yr | 77 | - | - | - | - |
| HHINC=3 | HH income $35,001 to $55,000/yr | 71 | - | - | - | - |
| HHINC=4 | HH income $55,001 to $75,000/yr | 48 | - | - | - | - |
| HHINC=5 | HH income $75,001 to $100,000/yr | 35 | - | - | - | - |
| HHINC=6 | HH income more than $100,000/yr | 39 | - | - | - | - |

* Observations in these household income dummies indicate counts for which income falls into this category.

## 5. FINDINGS

*Difference between groups*

The $CO_2$ emission levels in the experimental group are statistically significantly lower compared to the control group in Wave 2, while no significant differences exist between the two groups in Wave 1 (see Table 5). In Wave 2, the household-level daily $CO_2$ emission in the experimental group is 27.17 percent lower than that in the control group; the p-value of the two-sample t-test is 0.0457. For the households which did not change vehicle holdings (meaning the household had the same vehicles in Wave 1 and Wave 2), the difference in $CO_2$ emissions between the two groups is also statistically significant and even higher: the $CO_2$ emission from the experimental group is 32.6 percent lower than that from the control group when including only households that did not change vehicle holdings across the waves.

TABLE 5: Household-Level Personal Vehicle $CO_2$ Emission between Groups

|  |  | Wave 1 | | Wave 2 | |
| --- | --- | --- | --- | --- | --- |
|  |  | control | experimental | control | experimental |
| all households | Number of households with usable data | 80 | 80 | 80 | 80 |
|  | Number of vehicles in households with usable data | 99 | 89 | 93 | 91 |
|  | Average of daily household vehicle CO2 emissions (grams) | 9992.7 | 9371.1 | 10815.9 | 7877.5 |
|  | Difference: experimental-control) |  | -621.6 |  | -2938.4 |
|  | % Difference: (experimental-control)/control |  | -6.22% |  | -27.17% |
|  | p-value of two-sample t-test (two-sided) |  | 0.6527 |  | 0.0457 |
| households with no changes in vehicle holdings | Number of households with usable data | 62 | 61 | 62 | 61 |
|  | Number of vehicles in households with usable data | 69 | 66 | 69 | 66 |
|  | Average of daily household vehicle CO2 emissions (grams) | 9342.9 | 8814.3 | 11299.4 | 7615.6 |
|  | Difference: experimental-control) |  | -528.5 |  | -3683.8 |
|  | % Difference: (experimental-control)/control |  | -5.66% |  | -32.60% |
|  | p-value of two-sample t-test (two-sided) |  | 0.7300 |  | 0.0399 |

From Table 5, the daily average $CO_2$ emission of the households in the experimental group decreased from Wave 1 to Wave 2, while the emissions of the households in the control group increased between the two waves. The two-tailed p-value for the paired-sample t-test for Wave 1 to Wave 2 change in the experimental group is 0.0715, while the Wave 1 to Wave 2 change in the control group is not statistically significant (the paired t-test has a p-value of 0.4494.) There are also similar patterns of change for the households that did not change their vehicle holdings across these two waves. However, the Wave 1 to Wave 2 changes are not statistically significant for either the experimental or the control group subsets that had constant vehicle holdings, although we note the smaller sample size (n=62 control and n=61 experimental).

*Treatment effects by differences-in-differences estimator*

The set of estimation models using the DID estimator shows a statistically significant treatment effect of the opening of the Expo LRT line on household-level personal vehicle $CO_2$ emissions. Table 6 includes difference-in-difference models with different control variables. Without any socio- demographic covariates as controls, the treatment effect, as indicated as the coefficient of the wave-group interaction term, is not significant (Model 1). There are some treatment effects at a 10 percent significance level with person and vehicle counts controlled in Models 2 and 3. The most complete model, Model 4 which includes household income level dummy variables, shows a statistically significant treatment effect at the 5 percent level.

TABLE 6: Difference-in-Difference Estimation for Private Auto $CO_2$ Emission

|  | Model 1 | Model 2 | Model 3 | Model 4 |
|---|---|---|---|---|
| Experimental group dummy | -621.6 | 394.7 | 531.1 | 274.7 |
|  | (1419.3) | (1134.7) | (1120.8) | (1134.3) |
| Wave 2 dummy | 823.2 | 1433.0 | 1470.4 | 1342.0 |
|  | (1419.3) | (1133.1) | (1118.4) | (1124.5) |
| Wave experimental interaction | -2316.8 | -3129.8* | -3085.4* | -3145.0** |
|  | (2007.1) | (1602.3) | (1581.5) | (1588.1) |
| Number of household vehicles |  | 8130.1*** | 7282.6*** | 6152.1*** |
|  |  | (603.4) | (656.6) | (719.5) |
| Household size > 12 years old |  |  | 1616.0*** | 1794.6*** |
|  |  |  | (527.5) | (534.9) |
| Household annual income |  |  |  |  |
| $15,001 - $35,000 |  |  |  | 1466.1 |
|  |  |  |  | (1356.2) |
| $35,001 - $55,000 |  |  |  | 752.9 |
|  |  |  |  | (1385.9) |
| $55,001 - $75,000 |  |  |  | 3050.7** |
|  |  |  |  | (1528.1) |
| $75,001 - $100,000 |  |  |  | 5897.2*** |
|  |  |  |  | (1677.4) |
| $100,001 or more |  |  |  | 3760.5** |
|  |  |  |  | (1665.3) |
| Constant | 9992.7*** | -68.4 | -1827.3 | -2751.5* |
|  | (1003.6) | (1094.7) | (1223.5) | (1482.5) |
| N | 320 | 320 | 320 | 312 |
| adj. R-sq | 0.005 | 0.367 | 0.383 | 0.397 |

Standard errors in parentheses, * p<0.1, ** p<0.05, *** p<0.01

Model 4 is our preferred model because it controls for the number of persons and vehicles in the household and income levels and for household-specific changes in those values from Wave 1 to Wave 2. Household size, the number of vehicles, and income are all known to affect driving, and hence Model 4 isolates the treatment effect while controlling for any within-household changes in those values across the two waves. Model 4 shows a reduction in daily average household-level $CO_2$ emission of 3,145 grams from the Expo line opening, and given the research design we interpret that as a causal effect of the rail line on treatment group household private vehicle emissions relative to the control group. Note that the effect of the Expo line on $CO_2$ emissions among households living within ½ mile from the new stations is about half the size of the effect from adding an additional vehicle to the household (see Model 4). Given that vehicle holdings are strongly associated with both travel and private vehicle $CO_2$ emissions, the Expo line effect on nearby households is large in magnitude. Model 4 also shows that the $CO_2$ emission is higher with more vehicles and persons over 12 years old in the household and with higher household income levels, all of which are expected results.

*Travel Behavior Change*

Table 5 gives suggestive evidence that the bulk of the change in household $CO_2$ emissions is due to changes in driving behavior. Note that the pattern of emission changes is similar for the full sample and for the subset of households who did not change their vehicle holdings. Given that we are studying a short time window (from eight to three months before the Expo Line opened in Wave 1 to from five to eight months after opening in Wave 2), this is not surprising. Little time elapsed for households to purchase new vehicles and any impact of the Expo Line on vehicle holdings would likely be longer term. The data in Table 3 give insights into household travel behavior changes, and those values are the basis for the two-sample t-tests shown in Table 7. Average daily VMT per household in the experimental group is 6.81 miles lower than in the control group in Wave 2. For a more comprehensive analysis of travel changes from this data set, with results that give a pattern similar to Table 7, see Spears, Boarnet, and Houston (2015).

TABLE 7: Household VMT, Bus Trips and Train Trips

|  | Wave 1 | | Wave 2 | |
|---|---|---|---|---|
|  | **Control** | **Experimental** | **Control** | **Experimental** |
| **Household-level average daily VMT** | 25.60 | 24.29 | 27.56 | 20.75 |
| (standard deviation in parenthesis) | (2.26) | (2.83) | (3.13) | (1.98) |
| Number of households with usable data | 80 | 80 | 80 | 80 |
| Difference: experimental-control | | -1.30 | | -6.81 |
| p-value of two-sample t-test (two-sided) | | 0.7191 | | 0.0680 |
| **Household-level average daily bus trips** | 0.50 | 0.42 | 0.57 | 0.38 |
| (standard deviation in parenthesis) | (0.12) | (0.10) | (0.13) | (0.10) |
| Number of households with useable data | 78 | 78 | 78 | 78 |
| Difference: experimental-control | | -0.08 | | -0.19 |
| p-value of two-sample t-test (two-sided) | | 0.6230 | | 0.2378 |
| **Household-level average daily train trips** | 0.05 | 0.02 | 0.11 | 0.17 |
| (standard deviation in parenthesis) | (0.02) | (0.01) | (0.06) | (0.05) |
| Number of households with usable data | 78 | 78 | 78 | 78 |
| Difference: experimental-control | | -0.03 | | 0.06 |
| p-value of two-sample t-test (two-sided) | | 0.1539 | | 0.4165 |

## 6. LIFE-CYCLE CARBON EMISSION FROM TRANSIT

The previous section demonstrated that the opening of the Expo LRT line was significantly associated with a reduction in $CO_2$ emissions from personal vehicles. What about emission changes from changes in transit trip-making? Using the data in Table 7, and using paired t-tests to compare within-group changes (Wave 1 to Wave 2) in transit trips, bus trips do not change significantly for either group, while the experimental group households had a statistically significant increase in train trips (0.02 train trips per day in Wave 1, 0.17 train trips per day in Wave 2, p = 0.0006.) Life-cycle $CO_2$ emissions from changes in public transportation use for our sample can be estimated based on these travel behavior statistics.

There are two parts of the life-cycle $CO_2$ emissions associated with transit use: that from transit operation and that from the system building and maintenance. Specifically, carbon emission from transit operation includes the $CO_2$ generated from the consumption and production of the gas and electricity used by the buses and trains; and carbon emission from the system building and maintenance includes the $CO_2$ generated from (a) vehicle manufacturing and maintenance and (b) infrastructure (such as the light rail) construction and operation.

The first part (transit operation) of the life-cycle $CO_2$ emission can be estimated by calculating the average $CO_2$ emission per bus trip for LA Metro Buses and per train trip for LA Metro Rail. Based on transit operation data in the National Transit Database (NTD) and energy data from US EPA, Hodges (2010) estimated $CO_2$ emission (both pipeline and upstream) factor per passenger mile for the bus and rail transit facilities of the Los Angeles Metro. Specifically, Hodges (2010) estimated that average $CO_2$ emission associated with a passenger mile transit trip is 99.3 and 224.1 grams for light rail and bus, respectively, in the Los Angeles Metro system. We analyzed the NTD data and found that the average trip length for light rail and bus on the Los Angeles Metro system is 6.81 and 4.2 miles, respectively. Then we multiplied the per passenger mile factor from Hodges (2010) and the average trip lengths for LA Metro bus and Metro rail to obtain a per passenger trip emission factor for both bus and rail.[4] The factors are: (1) for LA Metro rail, one passenger trip is, on average, associated with 676.2 grams of $CO_2$ emission; (2)

---

[4] The travel survey that we administered asked respondent households to log the number of transit trips but did not ask them to record transit trip origins or destinations. Hence we cannot estimate the length of transit trips among our Expo line study subjects and thus we approximate transit trip lengths using the Los Angeles Metro system averages. This might overstate trip lengths, since the Expo Line is in a more central part of Los Angeles where trip distances might be shorter than in other parts of the Los Angeles Metro service area.

for LA Metro bus, one passenger trip is, on average, associated with 941.2 grams of $CO_2$ emission.

The second part (system building and maintenance) of the life-cycle $CO_2$ emission can be estimated by applying a scale factor. A recent study by Chester et al. (2013) estimated the short-term and long-term life-cycle environmental impacts (including $CO_2$ emissions) of the Los Angeles Metro Gold LRT and Orange Bus Rapid Transit (BRT) lines. The life-cycle $CO_2$ emissions from these two transit lines are not directly comparable to the Expo Line and the buses around our study area. However, since the two transit lines in the Chester et al. (2013) study are in the same transit system as the lines in our study, and lacking more specific data, we assume that the relative size of the $CO_2$ emission from this "system building and maintenance" part compared to the "transit operation" part can be used to estimate a scale factor.

Table 8 shows the calculation of the scale factor based on the study by Chester et al. (2013). As the table shows, the $CO_2$ emission associated with both lines can be divided into two parts: the transit operation part and the system building and maintenance part. The transit operation part includes $CO_2$ generated by: (1) vehicle operation (for buses), (2) propulsion electricity (for trains) and (3) energy production; while the system building and maintenance part includes $CO_2$ generated by (4) vehicle manufacturing and maintenance and (5) infrastructure construction and operation. Note that Chester et al. (2013) estimated separate $CO_2$ emission factors in the near-term (defined as to year 2030) and in the long-term (defined as 2030 – 2050). Since Chester et al. (2013) anticipated improvements in vehicle technology and cleaner electricity in the long-term, the life-cycle emission factor in the near-term (which is the value we use here) is higher than those in the long-term.

TABLE 8: Scale Factor Calculation for Life-Cycle $CO_2$ emission[5]

|  | Metro Gold LRT | Metro Orange BRT |
|---|---|---|
| Per passenger-mile $CO_2$ emission from: |  |  |
| (1) vehicle operation | 0 | 53.91 |
| (2) propulsion electricity | 120.73 | 0 |
| (3) energy production | 3.29 | 14.63 |
| (4) vehicle manufacturing and maintenance | 1.31 | 19.09 |
| (5) infrastructure construction and operation | 53.72 | 19.84 |
| (6) Per passenger-mile gross $CO_2$ emission (1-5) | 179.05 | 107.47 |
| (7) Per passenger-mile $CO_2$ emission associated with transit operation (1-3) | 124.02 | 68.54 |
| Scale factor (6)/(7) | 1.44 | 1.57 |

According to Table 8, the total life-cycle $CO_2$ emission (operation plus system building and maintenance) is 1.44 times the $CO_2$ emission from transit operations for the Metro Gold LRT; while the same factor for the Metro Orange BRT is 1.57. We use these scale factors to adjust the operational-only factors as follows: (1) for LA Metro rail, one passenger trip is, on average, associated with 973.7 grams of life-cycle $CO_2$ emission; (2) for LA Metro bus, one passenger trip is, on average, associated with 1477.7 grams of life-cycle $CO_2$ emission.

Based on these factors, the estimated household-level $CO_2$ emission from transit for the two groups across the two waves is shown in Table 9.

---

[5] Complete dataset of Chester et al. (2013) come from their online data base http://www.transportationlca.org/.

TABLE 9: Household-Level Transit-Related $CO_2$ Emission across Waves

|  | control | | experimental | |
|---|---|---|---|---|
|  | wave1 | wave2 | wave1 | wave2 |
| Number of households with usable data | 78 | 78 | 78 | 78 |
| Number of persons (12yrs or more) with usable data | 136 | 131 | 122 | 119 |
| Average of daily household-level CO2 emissions (grams) | 777.1 | 946.8 | 633.3 | 724.5 |
| Difference: wave2-wave1 | 169.7 | | 91.3 | |
| % Difference: (wave2-wave1)/wave1 | 21.8% | | 14.4% | |
| p-value of paired t-test (double-sided) | 0.1975 | | 0.3265 | |

From Table 9, transit-related household-level daily average life-cycle $CO_2$ emissions for the two groups both increased after the opening of the Expo Line: 21.8 percent increase for the control group and 14.4 percent increase for the experimental group. However, these changes are not statistically significant based on paired t-tests. The insignificance of the increases indicates that the change in $CO_2$ emission from transit could be omitted, and one could conclude that there is no countervailing increase in emissions from transit ridership. Of course, the sample size is small and the insignificant effect for transit emissions is likely due to the small sample size. Note, though, that the magnitude of the increase in total household daily transit emissions, 91.3 grams for the experimental group, is substantially smaller than the reduction in private vehicle emissions estimated in Model 4 of Table 6 (3145 grams) or the gap in the experimental group emissions versus control group emissions (2938 grams in Wave 2, from Table 5).

Alternatively, note that the change in life-cycle $CO_2$ emissions from transit combines an increase in light rail trips among the experimental group and a decrease in bus trips among the experimental group (see Table 7). Looking only at $CO_2$ emissions from light rail ridership, a two-sample t-test shows that light rail trips among the experimental group increased by 0.15 trips per day from Wave 1 to Wave 2, and the effect is statistically significant (t = 3.57). This implies that $CO_2$ emissions from light rail among the experimental group increased by 146 grams per day (0.15 more household LRT trips per day multiplied by 973.7 grams of $CO_2$ per light rail trip). That could be considered an upper bound of a countervailing transit emission effect, looking only at the statistically significant increase in light rail trips among the experimental group from Wave 1 to Wave 2. Even if we assume the insignificance of the paired t-tests in Table 9 comes from the relatively small sample size, the countervailing 146.0 grams per day of light rail emissions is much smaller than the reduction in private vehicle $CO_2$ emissions of 3,145 grams per day from our preferred DID model (Model 4, Table 6). On net, we conclude that the reduction in private vehicle emissions is, for all practical purposes, very close to the change in total transportation emissions that our research design attributes to our household sample as a result of the opening of the Expo LRT line.

## 7. DISCUSSION

*Sensitivity analysis*

This section tests the sensitivity of the models using different borders for the experimental area. An extensive literature in transportation and regional science uses ½ mile as the optimal light rail catchment distance. Guerra, Cervero, and Tischler (2012) tested different catchment distances, finding evidence that is consistent with the half-mile as the best distance for predicting transit ridership. Table 10 provides two additional DID models (Models 5 and 6) using 1km (0.62 mile) and 3/4 mile from home to station (straight-line) as break points between the experimental and control groups. In these two new models we included all control variables in Model 4 but only reported the wave variable, experimental – control group variable, and their interaction. The coefficients on the control variables in Models 5 and 6 are essentially unchanged, in sign, significance and magnitude from Model 4. According to Table 10, only Model 4 first introduced in the previous section has a significant treatment effect. Model 5 using 1km (0.62 mile) as break point between the experimental and control groups, has a treatment effect with similar magnitude as the one in Model 4 but is only significant at the 10 percent significance level. Model 6, using 3/4 mile as the dividing distance between experimental and control groups, does not have a statistically significant treatment effect on $CO_2$ emissions. We note that the small sample size of this study might influence these changes in significance across different experimental – control group distances, and we also note that the half-mile distance, long accepted in the literature, gives a statistically significant treatment effect on $CO_2$ emissions.

TABLE 10: Sensitivity Analysis of Vehicle $CO_2$ Emission DID Models

|  | Model 4 | Model 5 | Model 6 |
|---|---|---|---|
| Experimental group definition | < 1/2 mile | < 1km (0.62 mile) | < 3/4 mile |
| Experimental group dummy | 274.7 | 945.6 | -220.0 |
|  | (1134.3) | (1177.8) | (1214.0) |
| Wave 2 dummy | 1342.0 | 1684.8 | 1569.7 |
|  | (1124.5) | (1290.9) | (1376.4) |
| Wave experimental interaction | -3145.0** | -3088.9* | -2716.6 |
|  | (1588.1) | (1645.2) | (1686.8) |
| Total N | 312 | 312 | 312 |
| from experimental group | 157 | 193 | 208 |
| from control group | 155 | 119 | 104 |
| adj. R-sq | 0.397 | 0.392 | 0.396 |

Standard errors in parentheses, * $p<0.1$, ** $p<0.05$, *** $p<0.01$

*Rebound effects of new LRT service*

Economic theory suggests that the introduction of the Expo LRT line can impact the carbon emission for the households residing near the light rail station through two paths: first, by reducing the amount of driving by providing another travel mode choice – light rail – that would be preferred for at least some trips; second, by increasing the amount of driving if total travel costs are, on net, lower after the introduction of light rail.[6] The first effect is called the substitution effect and the second effect is called rebound effect (Small and Van Dender, 2007). The treatment effects we estimated in this paper are a combination of these two effects. Such "combined" treatment effects are effective from a program evaluation perspective. Given that we queried trip logs (counts of trips), we are not able to decompose travel change into a mode substitution and a rebound effect. In other work on the same sample data, Spears, Boarnet, and Houston (2015) argued that the largest cause of driving reductions among the experimental group was shorter driving trip lengths after the Expo LRT opened. This hints that travel changes are more complex than simple mode substitution and rebound effects. Yet we reiterate that our research design measures the full effect on VMT, without being able to decompose that effect into constituent travel behavior changes.

*Social costs*

While a quantitative cost-benefit analysis of the Expo LRT line is beyond the scope of this paper, it is worth mentioning some social costs related to the Line that may countervail the GHG reduction benefits measured here. The Measure R sales tax that helps fund construction of the Expo Line brings distortions. Small (1999) argues that sales tax transportation finance reduces the welfare of residents who are not planning on using the LRT Line.

---

[6] While for many trips, LRT can take more time than driving, we assume that increases in LRT use are due to generalized reductions in costs for trips, which can include convenience and the ability to avoid parking costs and peak hour traffic congestion. Given that we found evidence of VMT reductions in the experimental group after the Expo LRT opened (Table 7), it is reasonable to assume that for some persons in the sample, the Expo LRT is a lower cost mode for at least some trips. We note that our survey did not query trip costs or patterns of mode substitution and so we cannot measure differences in full travel costs across modes.

Another social cost is the potential for gentrification in the neighborhoods around the LRT stations. Theoretically, the housing units around the LRT line will become more expensive because of the increased accessibility, and the increased housing costs can displace long-time residents. Boarnet et al. (2015) discussed the severe housing affordability crisis in the Los Angeles region and documented increases in rents for new housing units near other rail transit lines in the Los Angeles Metro rail network. Having said that, a welfare analysis of the impacts on residents would be complex, having to account for the capitalization of improved accessibility into housing costs, differences in preferences across different groups of persons, and the welfare effect of residential moves into and out of the neighborhood.

Lastly, there is the question of whether a full accounting of social benefits would exceed the costs of building the LRT line, including the social costs. That is well beyond the scope of this paper. We address a different question, also vital for policy and regional science. Given the focus of many agencies worldwide on expanding rail transit as an approach to GHG reduction, there is still debate about the impact of such projects. Our program evaluation approach is intended to inform the narrower, but still important, question of how the Expo LRT line influenced changes in private vehicle GHG emissions.

## 8. CONCLUSION

A before-after, experimental-control group design is helpful in evaluating the effectiveness of transportation policy innovations. To date, this method has not been widely applied in transportation planning and policy evaluations. This study expanded the application of this method to analyze the effect of the opening of the Expo LRT line on personal vehicle $CO_2$ emissions among the households living within a half-mile of six new stations.

Through this analysis, we found that the opening of the LRT line is associated with lower daily average private vehicle $CO_2$ emissions for households near the stations. The opening of the Expo LRT line statistically significantly reduced the personal vehicle $CO_2$ emissions among the household living within ½ mile of the six selected stations from an average of 9,371.1 grams to 7,877.5 grams, a reduction of 15.94 percent; the opening of the line does not significantly impact the $CO_2$ emission levels on the households residing outside the ½ mile radii of these six stations. Our preferred DID model (Table 6, Model 4) indicates that the opening of the Expo LRT line is statistically significantly associated with a reduction of daily household private vehicle $CO_2$ emissions of 3,145 grams. When we redefine the experimental group to be households living within a kilometer of the new light rail stations, the DID estimate indicates emission reductions of 3,089 grams per day, although the effect is significant only at the ten percent level (Table 10.) Defining the experimental group as households living within ¾ mile from the new stations did not give a statistically significant emission reduction effect (Table 10), suggesting that the role of the new light rail on driving is larger for households living close (e.g. within ½ mile or a kilometer) of the new stations.

We note that in our study, as is typical of experimental designs, we can measure the full impact of the policy intervention on behavior, but we cannot decompose that impact into underlying structural components. In particular, our study design does not allow us to decompose the Expo Line effect on emissions into substitution and rebound effects in driving behavior, although we can measure the combined emission effect relative to a control group, which is a strength of the experimental research design. The results provide evidence that Los Angeles' rail transit investments can help reduce private vehicle GHG emissions. Beyond this specific case, we note that the experimental design used in this research can be applied to a broad range of transportation investments and policies, and we suggest continued application and refinement of before-after, experimental-control group study designs.

**ACKNOWLEDGEMENT**

The research was supported by the California Air Resources Board, the Haynes Foundation, the Lincoln Institute of Land Policy, the San Jose State Mineta Transportation Institute, the Southern California Association of Governments, the University of California Transportation Center, the University of California Multi-Campus Research Program on Sustainable Transportation, and the University of Southern California Lusk Center for Real Estate. The authors are grateful to the study participants. Gaby Abdel-Salam, Grecia Alberto, Priscilla Appiah, Gabriel Barreras, Sandip Chakrabarti, Gavin Ferguson, Lisa Frank, Dafne Gokcen, Andy Hong, Hsin-Ping Hsu, Jeongwoo Lee, Wei Li, Adrienne Lindgren, Greg Mayer, Carolina Sarmiento, Vicente Sauceda, Owen Serra, Steven Spears, Cynthia de la Torre, Dongwoo Yang, and Boyang Zhang assisted with data collection and processing at various points in the study. We also thank Mr. Nesamani Kalandiyur of California Air Resource Board for providing suggestions on the EMFAC model. We benefit from helpful comments from seminar participants at the June, 2014 symposium on "Low-Carbon Cites: Land Use and Transportation Intervention," at Shaanxi Normal University, Xi'an, China. The authors also acknowledge helpful comments and suggestions provided by coeditor Steven Brakman and two anonymous referees.

29